\documentclass[aps,pre,showpacs,amsmath,amssymb
,reprint,10pt,longbibliography]{revtex4-1}
\pdfoutput=1
\usepackage{bm}
\usepackage{graphicx}
\usepackage{xcolor}
\usepackage{newtxtext,newtxmath}
\usepackage{bbm}
\usepackage{mathrsfs} 
\usepackage{dsfont}
\usepackage{braket}

\usepackage[colorlinks=true,breaklinks=true,allcolors=blue]{hyperref}

\newcommand\CC{{\mathds{C}}}
\newcommand\id{{\mathds{1}}}
\newcommand{\mvec}[1]{\boldsymbol #1}

\DeclareMathOperator{\myIm}{\mathrm{Im}}
\DeclareMathOperator{\myRe}{\mathrm{Re}}

\makeatletter
\renewcommand{\p@paragraph}{}
\makeatother

\makeatletter
\let\cat@comma@active\@empty
\makeatother

\makeatletter
\g@addto@macro\bfseries{\boldmath}
\makeatother

\allowdisplaybreaks[1]

\begin{document}

\title{Ancilla-assisted probing of temporal quantum correlations of large spins}

\author{Michael Kastner} 
\affiliation{National Institute for Theoretical Physics (NITheP), Stellenbosch 7600, South Africa} 
\affiliation{Institute of Theoretical Physics, Department of Physics, University of Stellenbosch, Stellenbosch 7600, South Africa}

\date{\today}

\begin{abstract}
When measuring quantum spins at two or more different times, the later measurements are affected by measurement backaction occurring due to the earlier measurements. This makes the measurement of temporal quantum correlation functions challenging. In this paper, I propose a measurement protocol that mitigates the effect of measurement backaction by exploiting spin selection rules. I show that, under suitable conditions, the effect of measurement backaction on two-time quantum correlations becomes negligible when probing a system consisting of spins with large spin quantum numbers $l\gg s$ by coupling it to a spin-$s$ ancilla degree of freedom. A potential application of such a measurement protocol is the probing of an array of Bose--Einstein condensates by light.
\end{abstract}

\maketitle 

\section{Introduction}
\label{s:intro}
Temporal quantum correlations, such as the two-point function $\left\langle O_1(t_1)O_2(t_2)\right\rangle$, where $O_i(t_i)$ denotes the observable $O_i$ evolved in the Heisenberg picture until the time $t_i$, allow for a detailed characterization of the nonequilibrium behavior of quantum systems, more detailed than what equal-time correlations can achieve. Such correlation functions feature in a broad variety of physical theories and applications, including fluctuation--dissipation relations and the Kubo formula \cite{Kubo57}, scattering theory and optical coherences \cite{Glauber63}, transport theory \cite{Zwanzig65}, and glassy dynamics and aging \cite{SciollaPolettiKollath15}.  

Measuring temporal correlation functions is, in general, not an easy feat. A naive strategy might consist in measuring, at the time $t_1$, the expectation value of the operator $O_1O_2(t_2-t_1)$. However, at least in a many-body system with nontrivial time evolution, $O_2(t_2-t_1)$ is in general a nonlocal observable that is not directly experimentally accessible, and the product $O_1O_2(t_2-t_1)$ is likewise nonlocal and, in general, not even Hermitian. Another strategy for probing temporal quantum correlations might consist in measuring the observable $O_1$ at the early time $t_1$, followed by a measurement of the observable $O_2$ at the later time $t_2$. While in classical physics temporal correlations may indeed be obtained in this way, in quantum mechanics the measurement at the early time $t_1$ induces backaction on the system in the form of a wave-function collapse and, as a consequence, the outlined protocol does not yield the desired two-time quantum correlations.

Over the past decade, several protocols for the measurement of temporal quantum correlations that successfully mitigate or circumvent the above-described difficulties have been proposed \cite{RomeroIsart_etal12,Knap_etal13,Pedernales_etal14,GarciaAlvarez_etal17,Uhrich_etal17,KastnerUhrich18,UhrichGrossKastner19}. Most of these measurement protocols use ancilla quantum degrees of freedom onto which, by virtue of a system--ancilla interaction, information about the quantum system of interest is imprinted and then read out at a later stage. Each of these protocols comes with its own virtues as well as limitations, be it in the form of the resources required, the generality of applicability, or other. 

Here I introduce a measurement protocol that mitigates the effect of measurement backaction by exploiting spin selection rules. In this protocol, a spin-$s$ ancilla degree of freedom is coupled to the system at the time $t_1$ in a suitable way, followed by the measurement of $O_2$ at time $t_2$. I show that, under some additional conditions, this protocol yields the desired two-time correlation function if the local Hilbert space dimension of the constituents of the system of interest is much larger than $2s+1$. The intuition behind this proposal is that under these conditions, spin selection rules impose constraints on the effect of backaction. As an example, the reader may think of a chain of spin-$l$ degrees of freedom, probed by coupling a spin-$s$ to one of the spins of the chain through the unitary coupling operator $\exp(-i\lambda \mvec{S}_i\otimes \mvec{S})$, where $\mvec{S}$ denotes the spin vector operator of the ancilla, $\mvec{S}_i$ is the spin vector operator at the $i$th lattice site of the chain, and $\lambda$ is the coupling strength. The form of the coupling operator imposes selection rules and can facilitate only a restricted set of transitions between spin states, which are determined by the Clebsch--Gordan coefficients of the $\mvec{S}_i\otimes \mvec{S}$ coupled basis. Coupling to a spin-$s$ can change the spin-$l$ by, at most, $2s$, which is a small relative change if $s\ll l$, which in turn poses restrictions on the effect of backaction. This intuition will be made precise in Sec.~\ref{s:CandC}, where it is also shown that the constraint on the spin change implies a bound on the relative error of the temporal quantum correlation function measured by means of the ancilla-based measurement protocol.

The protocol I propose is similar to the measurement scheme of Refs.~\cite{Uhrich_etal17,KastnerUhrich18}, but differs in that it does not rely on a weak system--ancilla coupling. The benefit of dropping this weak-coupling requirement is that, compared to Refs.~\cite{Uhrich_etal17,KastnerUhrich18}, a significantly smaller number of repetitions of the measurement protocol is required in order to push statistical fluctuations below a desired threshold. Note that some of the existing measurement protocols for two-time quantum correlations, in particular those in Refs.\ \cite{RomeroIsart_etal12,Pedernales_etal14,GarciaAlvarez_etal17}, also use strong system--ancilla couplings, but are not based on spin selection rules and differ from the proposal of the present paper in several other aspects related to their generality, applicability, and performance. 

\begin{figure}\centering
  \includegraphics[width=0.96\linewidth]{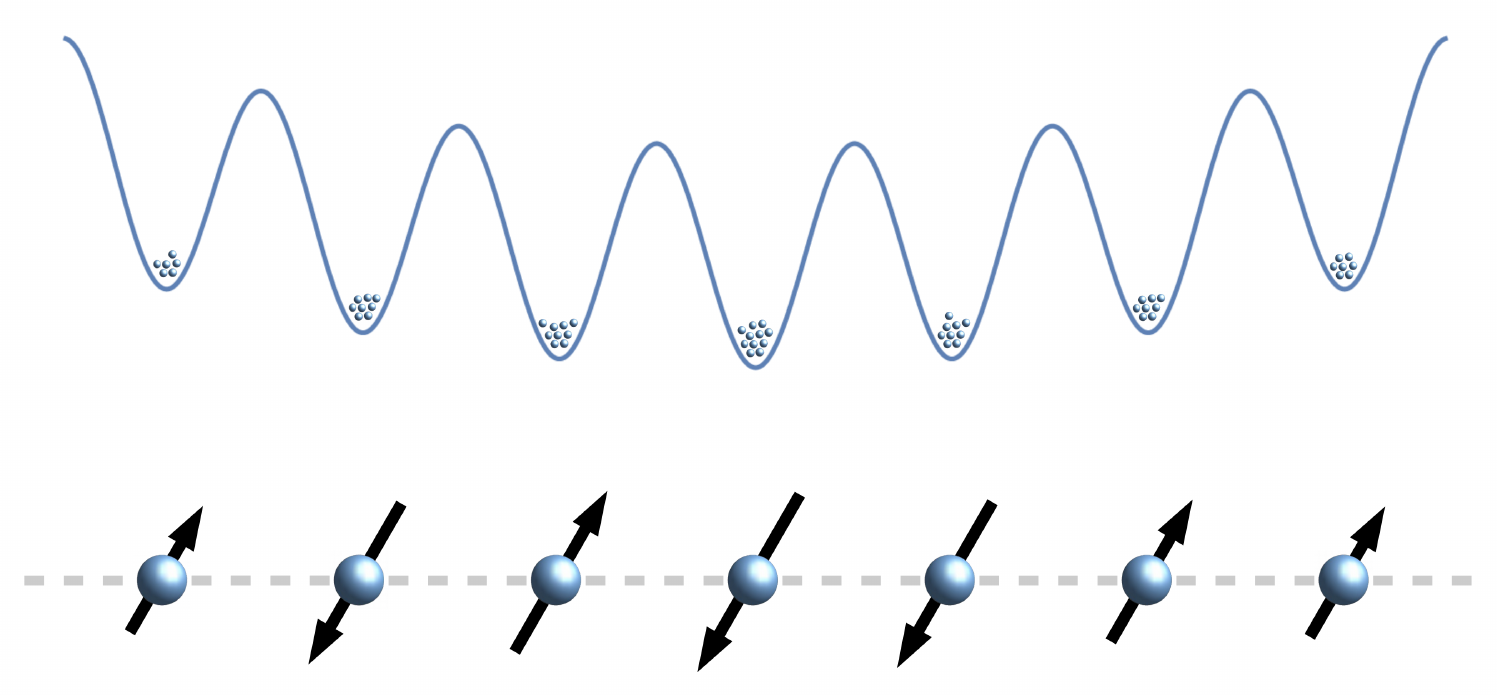}
	\caption{\label{f:System} 
		Top: A sinusoidal optical lattice, superimposed onto a parabolic trap potential, which holds a two-component Bose--Einstein condensate with (in general, site-dependent) atom number $N_i$ in each of its minima. Bottom: Each single condensate can be described by a pseudospin-$N_i/2$ degree of freedom at lattice site $i$, and neighboring spins can be made to interact (dashed gray lines) by adjusting the optical lattice depth.
	}
\end{figure}

The ancilla-based protocol of the present paper is, apart from the large-spin requirement, fairly versatile and may be applied to a number of settings and experimental platforms. The application I had in mind, and which motivated the development of the measurement protocol, is an optical lattice holding an array of two-component Bose--Einstein condensates, each of which consists of some tens or hundreds of atoms. In the parameter regime specified in Ref.~\cite{Muessel_etal14}, each single condensate can be described by a one-axis twisting Hamiltonian $H_i=\chi (J_i^z)^2$ \cite{KitagawaUeda93}, where $\chi$ denotes the coupling strength and $J_i^z$ is the $z$-component of a pseudospin-$N_i/2$ degree of freedom characterizing the $N_i$-atomic condensate at lattice site $i$. The atom numbers $N_i$ of the condensates, and hence also the spin quantum numbers $N_i/2$, can be determined experimentally with high accuracy \cite{Hume_etal13}. By adjusting the depth of the optical lattice, the pseudospins can be made to interact with each other; see Fig.~\ref{f:System} for an illustration. Following Refs.~\cite{KuzmichBigelowMandel98,KuzmichPolzik00}, a beam of off-resonant light shone onto the $i$th condensate induces an interaction proportional to $S^z J_i^z$, where $S^z$ denotes the $z$-component of the operator of the Stokes vector of the optical field integrated over the duration of the interaction. This operator representation of the Stokes vector is formally equivalent to a spin-$1/2$ operator, and hence the interaction $S^z J_i^z$ is equivalent to the coupling of a spin-$1/2$ to a a spin-$N_i/2$, where the number $N_i$ of atoms in the $i$th well is assumed to be much larger than 1.

\section{Setting}

I consider a spatially extended system consisting of spin degrees of freedom on a lattice of arbitrary size, structure, and lattice dimension. To each lattice site $i$, a spin-$l$ degree of freedom is assigned in the form of a local Hilbert space $\mathscr{H}=\CC^{2l+1}$ for all $i$. The total system consisting of $N$ spin-$l$ degrees of freedom has therefore the system Hilbert space $\mathscr{H}_\text{S}=\mathscr{H}^{\otimes N}$. Time evolution of the spin system is generated by an arbitrary system Hamiltonian $H$ acting on $\mathscr{H}_\text{S}$. (Generalization to nonunitary dynamics should be straightforward.)

The goal is to extract two-time correlation functions of the type
\begin{equation}\label{e:Ct1t2general}
C(t_1,t_2)=\langle O_1(t_1)O_2(t_2)\rangle,
\end{equation}
where $O_1$ and $O_2$ are arbitrary observables on $\mathscr{H}_\text{S}$. Typically, one may think of $O_1$ and $O_2$ as being supported on a single lattice site or on a few adjacent lattice sites. Such a restriction is not necessary for the theoretical aspects of the protocol developed in the following, but it is expected to greatly facilitate an experimental implementation. $O_i(t_1)=e^{i H t_1}O_1 e^{-i H t_1}$ denotes the observable $O_1$ that is time evolved in the Heisenberg picture under the Hamiltonian $H$ until time $t_1$.

\section{Measurement protocol}
\label{s:protocol}

For concreteness, I consider two-time correlations $C$ of the $z$-com\-po\-nents of spins at lattice sites $i$ and $j$ with respect to an arbitrary initial state $|\psi\rangle\in\mathscr{H}_\text{S}$,
\begin{equation}\label{e:Ct1t2}
\begin{split}
C(t_1,t_2)&=\langle S_i^z(t_1)S_j^z(t_2)\rangle\\
&= \bra{\psi} e^{i H t_1} S_{i}^{z}e^{-i H t_1}e^{i H t_2} S_j^{z}e^{-i H t_2}\ket{\psi},
\end{split}
\end{equation}
where $S_i^z$ denotes the $z$-component of the spin-$l$ operator acting on lattice site $i$. The restriction to $z$-components is only for notational convenience; results for other components are obtained by using angular momentum eigenstates with respect to $\mvec{S}^2$ and some other spin component in the derivation reported below. Generalizations to correlations at more than two times and/or more than two lattice sites should be feasible along similar lines.

I use a spin-$1/2$ degree of freedom as an ancilla, and hence $\mathscr{H}_\text{A}=\CC^2$ and the total Hilbert space is
\begin{equation}
\mathscr{H}_\text{tot} = \mathscr{H}_{\text{S}} \otimes \mathscr{H}_{\text{A}}.
\end{equation}
The system Hamiltonian $H=H_{\text{S}}\otimes\id_{\text{A}}$, which is responsible for the unitary evolution in the dynamic correlation function \eqref{e:Ct1t2}, acts nontrivially on $\mathscr{H}_{\text{S}}$ only. In the following, I outline a measurement protocol that permits one to determine a certain correlation function $\mathscr{C}$ defined below. Subsequently, in Sec.~\ref{s:CandC}, I show that the desired correlation function $C$ can be extracted from $\mathscr{C}$ under suitable conditions. The measurement protocol consists of the following steps.

\paragraph{Initial state preparation.}
Ancilla and system are assumed to be initially in a product state,
\begin{equation}\label{e:phi_psi}
\ket{\Psi}=\ket{\psi}\otimes\ket{\phi}\equiv\ket{\psi,\phi}.
\end{equation}
While the system initial state $\ket{\psi}$ is arbitrary (and determined by the physical situation under investigation), the ancilla initial state is prepared as
\begin{equation}\label{e:ancilla_ini}
\ket{\phi}=\frac{1}{\sqrt{2}}\left(\ket{-}+\ket{+}\right),
\end{equation}
where $|\pm\rangle$ denote the two eigenstates of the ancilla spin operator $S^z$ with corresponding eigenvalues $\pm 1/2$.

\paragraph{Time evolution until time $t_1$.}
Time-evolve the initial state $\ket{\Psi}$ up to the time $t_1$ with the system Hamiltonian $H_\text{S}$,
\begin{equation}
\ket{\Psi(t_1)}=e^{-i H_\text{S} t_1}\ket{\psi}\otimes\ket{\phi}\equiv\ket{\psi(t_1),\phi} .
\end{equation}
The ancilla state $\ket{\phi}$ remains unaffected.

\paragraph{Coupling of the ancilla to system site $i$.\label{para:couple}} 
Evolution of $\ket{\Psi(t_1)}$ with a unitary evolution operator
\begin{equation}\label{e:coupling}
 \mathscr{U}(\lambda) = \exp(-i\lambda \mvec{S}_i\otimes \mvec{S}),
\end{equation}
which typically induces entanglement between the system and the ancilla. At the end of the coupling procedure one obtains
\begin{equation}\label{e:post_coupling_state}
\ket{\Psi_\lambda(t_1)} = \mathscr{U}(\lambda)\ket{\psi(t_1),\phi}.
\end{equation}
Attempts to use other coupling operators, in particular $\exp(-i\lambda S_i^x\otimes S^x)$ and $\exp(-i\lambda S_i^z\otimes S^z)$, have been unsuccessful so far; see the Appendix \ref{a:zz} for details.

\paragraph{Measuring the ancilla.}\label{p:measuringancilla}
The ancilla is then probed by projectively measuring $\id_{\text{S}}\otimes S^z$, i.e., the $z$-component of the ancilla spin. According to the Born rule, one measures the eigenvalues $+1/2$ and $-1/2$ with probabilities
\begin{equation}\label{e:probt1}
	P_\pm = \left\langle\Psi_\lambda(t_1)\middle|\left(\id_{\text{S}}\otimes\ket{\pm}\bra{\pm} \right)\middle|\Psi_\lambda(t_1)\right\rangle.
\end{equation}
The post-measurement state is given by the normalized projection onto the subspace corresponding to the outcome $\pm1/2$ of the measurement,
\begin{equation}
\ket{\Psi_{\pm}(t_1)} = \frac{\left( \id_{\text{S}}\otimes \ket{\pm}\bra{\pm} \right)\ket{\Psi_\lambda(t_1)}}{\left\lVert\left( \id_{\text{S}}\otimes \ket{\pm}\bra{\pm} \right)\ket{\Psi_\lambda(t_1)}\right\rVert}\equiv\ket{\psi_\pm(t_1)}\otimes\ket{\pm}.
\end{equation}
Ancilla and system are again in a product state.

\paragraph{Time evolution until time $t_2$.}
Time-evolve the post-measurement state $\ket{\Psi_{\pm}(t_1)}$ up to the time $t_2$ with the system Hamiltonian $H_S$,
\begin{equation}
\ket{\Psi_{\pm}(t_2)} = e^{-i H_\text{S} (t_2-t_1)}\ket{\psi_{\pm}(t_1)}\otimes\ket{\pm}.
\end{equation}
The ancilla state $\ket{\pm}$ remains unaffected.

\paragraph{Projective measurement at site $j$.}
At the final time $t_2$, the disturbing effect due to a measurement is of no concern, and the observable $S_j^z$ at lattice site $j$ can be measured projectively without compromising the accuracy of the correlation function \eqref{e:Ct1t2} to be determined. I denote by $m \in \mathscr{S}$ the eigenvalues of $S_j^z$, where
\begin{equation}\label{e:Sset}
\mathscr{S} = \lbrace -l,-l+1, \dotsc, l-1,l\rbrace.
\end{equation}
The conditional probability of obtaining $m$ when measuring $S_j^b$, after having obtained $\pm1/2$ when measuring the ancilla, is
\begin{equation}\label{e:conditional_probt2}
P_{m\vert \pm} = \bigl\langle\Psi_\pm(t_2) \big| (\ket{m}\bra{m}\otimes\mathds{1}_{\text{A}}) \big| \Psi_{\pm}(t_2)\bigr\rangle.
\end{equation}

\paragraph{Correlating the measured outcomes.}
\label{p:g}
The probabilities \eqref{e:probt1} and \eqref{e:conditional_probt2} are used to calculate the correlation between the measured ancilla spin at $t_1$ and the system spin $j$ at $t_2$,
\begin{align}\label{e:wmcf}
\mathscr{C}(t_1,t_2) &= \sum_{m\in \mathscr{S} } m \left(P_{m \vert +} P_{+} - P_{m \vert -} P_{-}\right)\\
&=\Braket{\Psi(t_1)|\mathscr{U}^\dagger(\lambda)\left(S_j^z(t_2-t_1)\otimes S^z\right) \mathscr{U}(\lambda)|\Psi(t_1)},\nonumber
\end{align}
where the second line has been obtained by using the definitions introduced in this section and by making use of the spectral representations of $S_j^z$ and of the ancilla spin $S^z$.

\section{Relating \texorpdfstring{$\mathscr{C}$}{\mathscr{C}} to \texorpdfstring{$C$}{C}}
\label{s:CandC}

In this section, I show that the correlation function $\mathscr{C}(t_1,t_2)$ defined in Eq.~\eqref{e:wmcf} contains, under suitable conditions, terms proportional to the real, respectively imaginary, part of the desired two-time correlation function $C(t_1,t_2)$. A reader not interested in the details of the derivation may skip most of this section and continue reading from the main result \eqref{e:C_final}. The assumptions made in the course of the derivation are summarized in Sec.~\ref{s:summary}.

The main tool for the calculation is to make use of the angular momentum-coupled basis when dealing with the spin--spin coupling operator $\mathscr{U}(\lambda)$ in Eq.~\eqref{e:coupling}, and to resort to the uncoupled basis when evaluating the effect of the ancilla spin operator $S^z$ and when exploiting product properties of the initial state. The switching between the two bases is facilitated by means of Clebsch--Gordan coefficients.

The joint system--ancilla Hilbert space is written as a tensor product space consisting of three factors,
\begin{equation}
\mathscr{H}_\text{tot}=\mathscr{H}_i\otimes\mathscr{H}_\text{A}\otimes\mathscr{H}_\text{rest},
\end{equation} 
where the Hilbert spaces $\mathscr{H}_i$ and $\mathscr{H}_\text{A}$ of lattice site $i$ and the ancilla, respectively, are treated separately from the rest of the system. I introduce a coupled basis $\ket{l,j,m}$ on the subspace $\mathscr{H}_i\otimes\mathscr{H}_\text{A}$, defined through the following eigenvalue equations,
\begin{subequations}
\begin{align}
\mvec{S}_i^2\ket{l,j,m}&=l(l+1)\ket{l,j,m},\\
\mvec{S}^2\ket{l,j,m}&=(3/4)\ket{l,j,m},\\
\mvec{J}^2\ket{l,j,m}&=j(j+1)\ket{l,j,m},\\
J^z\ket{l,j,m}&=m\ket{l,j,m},
\end{align}
\end{subequations}
with $\mvec{J}=\mvec{S}_i+\mvec{S}$, where $j\in\{l-1/2,l+1/2\}$ and $m\in\{-j,-j+1,\dotsc,j\}$ (and similarly for $\tilde\jmath$ and $\tilde m$ introduced below). The coupled basis is particularly useful for evaluating the system--ancilla coupling unitary,
\begin{multline}
\mathscr{U}(\lambda)\ket{l,j,m}=\exp\bigl[-i\lambda\bigl(\mvec{J}^2-\mvec{S}_i^2-\mvec{S}^2\bigr)/2\bigr]\ket{l,j,m}\\
=\exp\Bigl\{-i\lambda\bigl[j(j+1)-l(l+1)-\tfrac{3}{4}\bigr]/2\Bigr\}\ket{l,j,m}.
\end{multline}

To make use of this diagonal form of the coupling unitary, the pre-coupling state at time $t_1$ is expanded in terms of the coupled basis,
\begin{equation}\label{e:expansion}
\ket{\Psi(t_1)}=\sum_{j,m,\mvec\alpha}c_{jm}\ket{l,j,m;\mvec\alpha},
\end{equation}
where $c_{jm}$ are complex-valued expansion coefficients and $\mvec\alpha$ denotes a set of quantum numbers labeling some basis of $\mathscr{H}_\text{rest}$. By inserting this expansion into the correlation function \eqref{e:wmcf} one obtains
\begin{multline}\label{e:C}
\mathscr{C}=\sum_{j,m,\mvec\alpha}\sum_{\tilde{\jmath},\tilde{m},\tilde{\mvec\alpha}}c_{\tilde{\jmath}\tilde{m}\tilde{\mvec\alpha}}^* c_{jm\mvec\alpha}\varepsilon_{j\tilde{\jmath}}(\lambda)\\
\times\Braket{l,\tilde{\jmath},\tilde{m};\tilde{\mvec\alpha}|S_j^z(t_2-t_1)\otimes S^z|l,j,m;\mvec\alpha}
\end{multline}
with
\begin{equation}
\varepsilon_{j\tilde{\jmath}}(\lambda):=\exp\left\{\frac{i\lambda}{2}\left[\tilde\jmath(\tilde\jmath+1)-j(j+1)\right]\right\}.
\end{equation}
To shorten the notation, I will, in the following, omit the $\mvec\alpha$ quantum numbers and write
\begin{equation}\label{e:C_v1}
\mathscr{C}=\sum_{j,m}\sum_{\tilde{\jmath},\tilde{m}}c_{\tilde{\jmath}\tilde{m}}^* c_{jm}\varepsilon_{j\tilde{\jmath}}(\lambda) \Braket{l,\tilde{\jmath},\tilde{m}|S_j^z(t_2-t_1)\otimes S^z|l,j,m}.
\end{equation}
However, the reader should bear in mind that all results hold for very general many-body systems, and not only for a single large spin coupled to a spin-$1/2$ ancilla.

The transformation to the uncoupled basis is, for the case of an arbitrary angular momentum $l$ coupled to a spin-$1/2$, given by \cite[Chap.~15.2]{Shankar}
\begin{equation}\label{e:ClebschGordan}
\Ket{l,l\pm\tfrac{1}{2},m}=a_{lm}\Ket{l,m\mp\tfrac{1}{2},\tfrac{1}{2},\pm\tfrac{1}{2}}\pm b_{lm}\Ket{l,m\pm\tfrac{1}{2},\tfrac{1}{2},\mp\tfrac{1}{2}}
\end{equation}
for all $|m|\leqslant l\pm1/2$, with Clebsch--Gordan coefficients
\begin{equation}\label{e:CG_coefficients}
a_{lm}=\sqrt{\frac{l+1/2+m}{2l+1}},\qquad b_{lm}=\sqrt{\frac{l+1/2-m}{2l+1}}.
\end{equation}
Here, kets with four entries (compared to three entries for the coupled basis) denote the elements of the uncoupled basis, which satisfy the eigenvalue equations
\begin{subequations}
\begin{align}
\mvec{S}_i^2\ket{l,m_l,s,m_s}&=l(l+1)\ket{l,m_l,s,m_s},\\
S_i^z\ket{l,m_l,s,m_s}&=m_l\ket{l,m_l,s,m_s},\\
\mvec{S}^2\ket{l,m_l,s,m_s}&=s(s+1)\ket{l,m_l,s,m_s},\\
S^z\ket{l,m_l,s,m_s}&=m_s\ket{l,m_l,s,m_s}.\label{e:Sz}
\end{align}
\end{subequations}
By inserting \eqref{e:ClebschGordan} into \eqref{e:C_v1} and using \eqref{e:Sz} as well as the orthonormality of the basis states, one obtains
\begin{widetext}
\begin{align}\label{e:C_v2}
\mathscr{C}=\tfrac{1}{2}\sum_{m,\tilde m}&\bigg\{\Braket{l,\tilde m-\tfrac{1}{2}|S_j^z(t_2-t_1)|l,m-\tfrac{1}{2}}\\
&\qquad\times\left[c_{+,\tilde m}^* c_{+,m}a_{l\tilde m}a_{lm}+c_{-,\tilde m}^* c_{-,m}b_{l\tilde m}b_{lm}-c_{+,\tilde m}^* c_{-,m}a_{l\tilde m}b_{lm}e^{i\lambda(l+1/2)}-c_{-,\tilde m}^* c_{+,m}b_{l\tilde m}a_{lm}e^{-i\lambda(l+1/2)} \right]\nonumber\\
&-\Braket{l,\tilde m+\tfrac{1}{2}|S_j^z(t_2-t_1)|l,m+\tfrac{1}{2}}\nonumber\\
&\qquad\times\left[c_{+,\tilde m}^* c_{+,m}b_{l\tilde m}b_{lm}+c_{-,\tilde m}^* c_{-,m}a_{l\tilde m}a_{lm}+c_{+,\tilde m}^* c_{-,m}b_{l\tilde m}a_{lm}e^{i\lambda(l+1/2)}+c_{-,\tilde m}^* c_{+,m}a_{l\tilde m}b_{lm}e^{-i\lambda(l+1/2)} \right]\bigg\},\nonumber
\end{align} 
\end{widetext}
where the shorthand $c_{\pm,m}\equiv c_{l\pm1/2,m}$ has been used. The ancilla-part of the expectation values has already been evaluated, and the remaining matrix elements in \eqref{e:C_v2} involve only operators on and states from the system Hilbert space $\mathscr{H}_\text{S}=\mathscr{H}_i\otimes\mathscr{H}_\text{rest}$. (Recall that the $\mvec\alpha$ quantum numbers referring to $\mathscr{H}_\text{rest}$ have been omitted, and the matrix elements in \eqref{e:C_v2} really involve many-body states and operators.)

The ancilla initial state, as well as the system state at time $t_1$, are encoded in the expansion coefficients $c_{\pm,m}$, which, as is evident from \eqref{e:expansion}, refer to the coupled basis. To make use of the properties of the ancilla initial state \eqref{e:ancilla_ini}, these coefficients need to be translated into expansion coefficients of the uncoupled basis. Making use of \eqref{e:expansion} and \eqref{e:ClebschGordan}, one obtains
\begin{equation}\label{e:c_expansion}
c_{\pm,m}=a_{lm}\gamma_m^\pm \pm b_{lm}\gamma_m^\mp,
\end{equation}
where
\begin{equation}\label{e:gamma}
\gamma_m^\pm = \Braket{l,m\mp\tfrac{1}{2},\tfrac{1}{2},\pm\tfrac{1}{2}|\Psi(t_1)}
\end{equation}
denote the expansion coefficients of $\ket{\Psi(t_1)}$ with respect to the uncoupled basis. The form of the ancilla initial state \eqref{e:ancilla_ini} implies that
\begin{equation}
\Braket{l,m,\tfrac{1}{2},+\tfrac{1}{2}|\Psi(t_1)}=\Braket{l,m,\tfrac{1}{2},-\tfrac{1}{2}|\Psi(t_1)},
\end{equation}
and hence $\gamma_{m+1}^+=\gamma_m^-$. By inserting \eqref{e:c_expansion} into \eqref{e:C_v2}, and furthermore assuming $\gamma_{m+1}^\pm\approx\gamma_m^\pm\equiv\gamma_m$ (i.e., the expansion coefficients vary slowly with respect to $m$), one obtains
\begin{widetext}
\begin{align}\label{e:C_v3}
\mathscr{C}\approx\tfrac{1}{2}\sum_{m,\tilde m}\gamma_{\tilde m}^*\gamma_m&\bigg\{\Braket{l,\tilde m-\tfrac{1}{2}|S_j^z(t_2-t_1)|l,m-\tfrac{1}{2}}\\
&\qquad\times\left[(\tilde a+\tilde b)(a+b)\tilde a a + (\tilde a-\tilde b)(a-b)\tilde b b-(\tilde a+\tilde b)(a-b)\tilde a b e^{i\lambda(l+1/2)} - (\tilde a-\tilde b)(a+b)\tilde b a e^{-i\lambda(l+1/2)} \right]\nonumber\\
&-\Braket{l,\tilde m+\tfrac{1}{2}|S_j^z(t_2-t_1)|l,m+\tfrac{1}{2}}\nonumber\\
&\qquad\times\left[ (\tilde a+\tilde b)(a+b)\tilde b b + (\tilde a-\tilde b)(a-b)\tilde a a+(\tilde a+\tilde b)(a-b)\tilde b a e^{i\lambda(l+1/2)} + (\tilde a-\tilde b)(a+b)\tilde a b e^{-i\lambda(l+1/2)} \right]\bigg\},\nonumber
\end{align} 
\end{widetext}
where the shorthand notations $a\equiv a_{lm}$, $\tilde a\equiv a_{l\tilde m}$, $b\equiv b_{lm}$, and $\tilde b\equiv b_{l\tilde m}$ have been used. I further assume that the matrix elements in \eqref{e:C_v3} vary slowly with $m$, such that
\begin{equation}\label{e:matrix_approx}
\Braket{l,\tilde m\pm\tfrac{1}{2}|S_j^z(t_2-t_1)|l,m\pm\tfrac{1}{2}}\approx\Braket{l,\tilde m|S_j^z(t_2-t_1)|l,m}.
\end{equation}
Under this condition, \eqref{e:C_v3} simplifies to
\begin{multline}\label{e:C_v4}
\mathscr{C}\approx \sum_{m,\tilde m}\gamma_{\tilde m}^*\gamma_m\Braket{l,\tilde m|S_j^z(t_2-t_1)|l,m} (\tilde a b+a\tilde b)\\
\times\left\{(a\tilde a-b\tilde b)[1-\cos(\lambda l)]+(\tilde a b-a\tilde b)i\sin(\lambda l)\right\},
\end{multline}
where I have approximated $l+1/2\approx l$ in the trigonometric functions \footnote{This approximation is inessential and is only introduced to shorten the notation. It can be undone by replacing $l$ with $l+1/2$ in the trigonometric functions of the final result \eqref{e:C_final}.}. Using the definitions of the Clebsch--Gordan coefficients \eqref{e:CG_coefficients} and approximating their denominators by $2l+1\approx2l$, the correlation function can be rewritten as
\begin{multline}\label{e:C_v5}
\mathscr{C}\approx\frac{1}{2l}\sum_{m,\tilde m}\gamma_{\tilde m}^*\gamma_m\Braket{l,\tilde m|S_j^z(t_2-t_1)|l,m} \Biggl\{(\tilde m-m)i\sin(\lambda l)\\
+2\left(\tilde m\sqrt{1-(m/l)^2}+m\sqrt{1-(\tilde m/l)^2}\right)\sin^2(\lambda l/2)\Biggr\}.
\end{multline}
Making use of the spectral theorem, one can write
\begin{widetext}
\begin{multline}\label{e:C_v6}
\mathscr{C}\approx\frac{1}{2l}\sum_{m,\tilde m}\gamma_{\tilde m}^*\gamma_m\biggl\{\Braket{l,\tilde m|\left[S_i^z,S_j^z(t_2-t_1)\right]|l,m} i\sin(\lambda l)\\*
+2\Braket{l,\tilde m|S_i^z S_j^z(t_2-t_1)\sqrt{1-(S_i^z/\mvec{S}_i)^2}+\sqrt{1-(S_i^z/\mvec{S}_i)^2}S_j^z(t_2-t_1)S_i^z|l,m}\sin^2(\lambda l/2)\biggr\}\\
=\frac{1}{2l}\Braket{\Psi(t_1)|\left(S_i^z,S_j^z(t_2-t_1)\right)i\sin(\lambda l)+2\left(S_i^z S_j^z(t_2-t_1)\sqrt{1-(S_i^z/\mvec{S}_i)^2}+\sqrt{1-(S_i^z/\mvec{S}_i)^2}S_j^z(t_2-t_1)S_i^z\right)\sin^2(\lambda l/2)|\Psi(t_1)}.
\end{multline}
\end{widetext}
Assuming that $\sqrt{1-(S_i^z/\mvec{S}_i)^2}\approx1$
, the correlation function simplifies to
\begin{multline}\label{e:C_v7}
\mathscr{C}\approx\frac{1}{l}\Braket{\Psi(t_1)|\left\{S_i^z,S_j^z(t_2-t_1)\right\}|\Psi(t_1)}\sin^2(\lambda l/2)\\
+\frac{i}{2l}\Braket{\Psi(t_1)|\left[S_i^z,S_j^z(t_2-t_1)\right]|\Psi(t_1)}\sin(\lambda l),
\end{multline}
where the curly brackets denote the anticommutator. Using the definition of $\ket{\Psi(t_1)}$, this can be cast into the form
\begin{multline}\label{e:C_v8}
\mathscr{C}\approx\frac{2}{l}\myRe\Braket{\Psi|S_i^z(t_1) S_j^z(t_2)|\Psi}\sin^2(\lambda l/2)\\
+\frac{1}{l}\myIm\Braket{\Psi|S_i^z(t_1) S_j^z(t_2)|\Psi}\sin(\lambda l)
\end{multline}
or, equivalently,
\begin{equation}\label{e:C_final}
\mathscr{C}\approx\frac{2}{l}\sin^2(\lambda l/2)\myRe C(t_1,t_2)+\frac{1}{l}\sin(\lambda l)\myIm C(t_1,t_2),
\end{equation}
which is the main result of this paper. Evidently, $\mathscr{C}$ contains both the real and imaginary parts of the desired two-time correlation function $C$ defined in \eqref{e:Ct1t2}. A possible strategy for separating the real part from the imaginary part consists in measuring, according to the protocol of Sec.~\ref{s:protocol}, the correlation function $\mathscr{C}$ at different coupling times/strength $\lambda$. For example, choosing $\lambda l=\pi$ gives
\begin{equation}\label{e:ReC}
\mathscr{C}(\lambda l=\pi)\approx\frac{2}{l}\myRe C(t_1,t_2),
\end{equation}
from which an estimate of $\myRe C$ can be obtained. By further measuring an estimate of $\mathscr{C}$ at $\lambda l=\pi/2$,
\begin{equation}
\mathscr{C}(\lambda l=\pi/2)\approx\frac{1}{l}\myRe C(t_1,t_2)+\frac{1}{l}\myIm C(t_1,t_2),
\end{equation}
and making use of the knowledge of $\myRe C$, the imaginary part $\myIm C$ can be extracted. More generally, Fourier analysis can be used to extract real and imaginary parts of $C$ from estimators of $\mathscr{C}$ at multiple and arbitrarily spaced values of $\lambda l$.


\section{Example: two spin-\texorpdfstring{$l$}{l} coupled to spin-\texorpdfstring{$1/2$}{1/2}}
\label{s:example}

To illustrate the performance of the protocol of Sec.~\ref{s:protocol}, and also its statistical and systematic errors, I study a simple system consisting of two coupled spin-$l$ degrees of freedom, augmented by a spin-$1/2$ ancilla used for probing at the early time $t_1$. The total Hilbert space is
\begin{equation}
\mathscr{H}_\text{tot}=\mathscr{H}_1 \otimes \mathscr{H}_2 \otimes \mathscr{H}_\text{A}
\end{equation}
with $\mathscr{H}_1=\CC^{2l+1}=\mathscr{H}_2$ and $\mathscr{H}_\text{A}=\CC^2$. 
For the system Hamiltonian I choose a Heisenberg coupling between the two spin-$l$ degrees of freedom,
\begin{equation}
H_\text{S}=\mvec{S}_1 \otimes \mvec{S}_2 \otimes \id_\text{A}.
\end{equation}
The goal is to measure the normalized two-time correlation function
\begin{equation}\label{e:C12}
C(t_1,t_2)=\frac{\langle S_1^z(t_1)S_2^z(t_2)\rangle}{l^2}
\end{equation}
with respect to initial states specified further below. The ancilla initial state and the system--ancilla coupling are as specified in Eqs.~\eqref{e:ancilla_ini} and \eqref{e:coupling}. To obtain the correlation function $\mathscr{C}$ in Eq.~\eqref{e:wmcf}, one needs to calculate the probabilities \eqref{e:probt1} and \eqref{e:conditional_probt2}, which are given as expectation values of certain projection operators. For the three-spin example considered here, these expectation values can be numerically computed in Mathematica for moderate spin quantum numbers $l$.

\begin{figure}\centering
\includegraphics[width=0.5\linewidth]{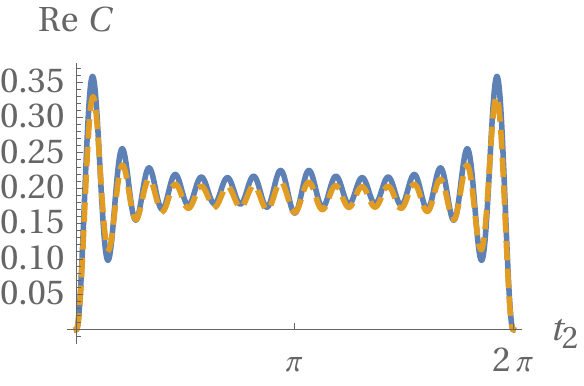}
\hfill
\includegraphics[width=0.47\linewidth]{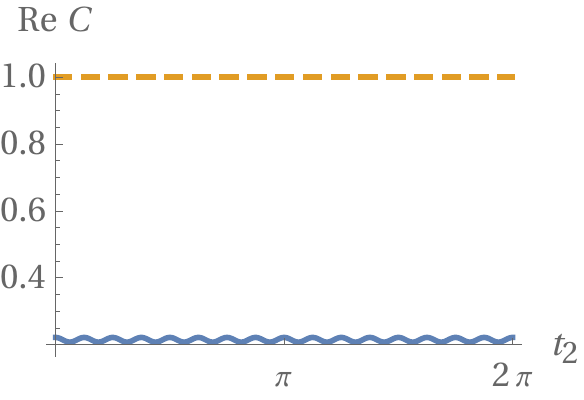}
\caption{\label{f:examples} 
Real parts of two-time correlation functions $C(t_1,t_2)$ as defined in \eqref{e:C12} for spin quantum number $l=8$, plotted for $t_1=0$ as a function of $t_2$. Orange dashed lines show exact correlation functions, solid blue lines are estimates based on Eqs.~\eqref{e:ReC} and \eqref{e:wmcf}. Left: Starting from the uniform initial state \eqref{e:Psiini_uniform}, which satisfies the condition of slowly varying expansion coefficients $\gamma_m^\pm$, good agreement between exact and estimated correlations is observed. Right: For the maximally magnetized initial state \eqref{e:psiini_max}, the coefficients $\gamma_m^\pm$ are not slowly varying in $m$, and hence the conditions for the validity of the measurement protocol are not satisfied, leading to a large discrepancy between exact and estimated correlations.}
\end{figure}

For this setting, I show in Fig.~\ref{f:examples} the real part of the exact correlation function \eqref{e:C12}, and compare it to the estimate calculated via \eqref{e:ReC} and \eqref{e:wmcf}. The left plot in Fig.~\ref{f:examples} is for a system initial state
\begin{equation}\label{e:Psiini_uniform}
\frac{1}{2l+1}\sum_{m_1,m_2=-l}^l\ket{l,m_1}\otimes\ket{l,m_2},
\end{equation}
in which all $S_i^z$ eigenstates are equally populated, and as a result the expansion coefficients $\gamma_m^\pm$ vary slowly with $m$, as required for the measurement protocol. Indeed, Fig.~\ref{f:examples} (left) shows good agreement between the exact two-time correlation function and its measured approximation. In contrast, the system initial state
\begin{equation}\label{e:psiini_max}
\ket{l,l}\otimes\ket{l,l},
\end{equation}
which has rapidly varying expansion coefficients $\gamma_m^\pm$, leads to a substantial discrepancy between the exact correlations and their measured counterpart (Fig.~\ref{f:examples}, right).

\begin{figure}\centering
\includegraphics[width=0.48\linewidth]{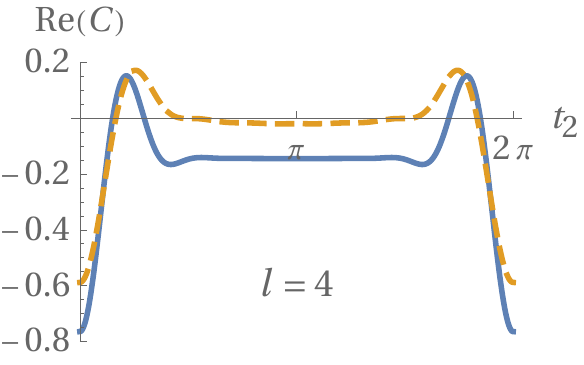}
\hfill
\includegraphics[width=0.48\linewidth]{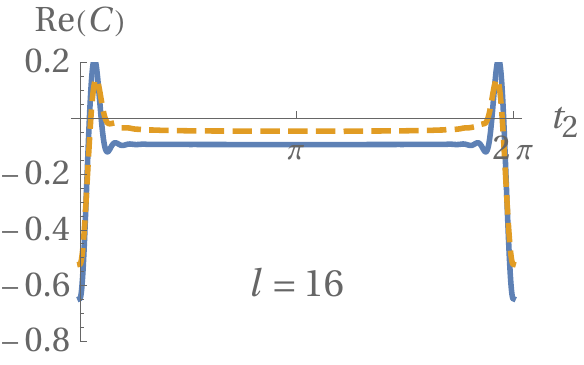}
\caption{\label{f:lscaling} 
Real parts of two-time correlation functions $C(t_1,t_2)$ as defined in \eqref{e:C12} starting from the initial state \eqref{e:Psiini_ramp}, plotted for $t_1=0$ as a function of $t_2$. Orange dashed lines show the exact correlation functions, blue solid lines are estimates based on Eqs.~\eqref{e:ReC} and \eqref{e:wmcf}, which correspond to the outcome of the measurement protocol of Sec.~\ref{s:protocol} in the limit of a large number of measurement runs. Hence, in the absence of statistical errors from finite sample sizes, the differences between the orange dashed and the blue solid lines illustrate the systematic error stemming from large-$l$ expansions and approximations in the derivation of the measurement protocol. For $l=4$ (left panel), the systematic error is significant, but shrinks rapidly with increasing spin quantum numbers $l=16$ (right panel).
}
\end{figure}

In addition to slowly varying $\gamma_m^\pm$, the protocol of Sec.~\ref{s:protocol} also requires a large spin quantum number $l$, and estimators of the two-time correlation function $C$ are expected to be more accurate the larger $l$ is. To explore how the accuracy of the measured estimator depends on the spin quantum number $l$, I use as a system initial state a normalized version of
\begin{equation}\label{e:Psiini_ramp}
\sum_{m_1=-l}^l (l-m_1)\ket{l,m_1}\otimes\ket{l,l}
\end{equation}
and calculate correlations for various spin quantum numbers $l$. For this choice, the measured estimator deviates significantly from the exact correlation function for $l=4$ (Fig.~\ref{f:lscaling} left), but the agreement improves rapidly when increasing the spin quantum number to $l=16$ (Fig.~\ref{f:lscaling} right).

The plots in Figs.~\ref{f:examples} and \ref{f:lscaling} are based on Eqs.~\eqref{e:ReC} and \eqref{e:wmcf}, making use of the probabilities $P_\pm P_{m|\pm}$ calculated according to Eqs.~\eqref{e:probt1} and \eqref{e:conditional_probt2}. In an experimental realization of the measurement protocol of Sec.~\ref{s:protocol}, however, the exact probability distributions are not available, but have to be estimated as a sample average over repeated runs of the protocol. To obtain the exact probabilities $P_\pm P_{m|\pm}$ in principle requires an infinite sample of runs and is not realistic. Finite samples, on the other hand, introduce errors in $P_\pm$ and $P_{m|\pm}$, which by error propagation result in statistical errors in the correlation function $C$, on top of the systematic errors discussed above and illustrated in Figs.~\ref{f:examples} and \ref{f:lscaling}. To assess the magnitude of these statistical errors, I proceed as follows: From the probability distribution $P_\pm P_{m|\pm}$ I draw samples of size $\mathscr{N}_\text{s}$. Each element of the sample represents the outcome $(\pm1/2,m)$ of a hypothetical measurement according to the protocol of Sec.~\ref{s:protocol}. To mimic experimental constraints I use, instead of the exact probabilities $P_\pm P_{m|\pm}$, the relative frequencies of the outcomes $(\pm1/2,m)$ within a random sample. The larger the sample size, the smaller is the statistical error in the probabilities, and hence also in the correlation function $C$. To estimate the size of these statistical errors for a given sample size $\mathscr{N}_\text{s}$, I calculate $C$ not only from one sample, but for 100 samples of size $\mathscr{N}_\text{s}$. Each of those samples will give a slightly different value of $C$, and I calculate the standard deviation of these fluctuating values, which serves as an estimate of the statistical error in $C$.

\begin{figure}\centering
\includegraphics[width=0.48\linewidth]{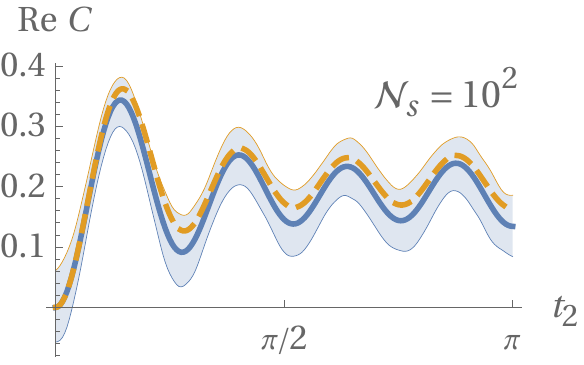}
\hfill
\includegraphics[width=0.48\linewidth]{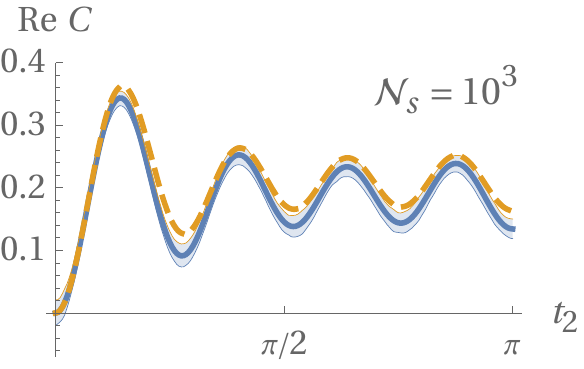}
\caption{\label{f:errors} 
Illustration of the magnitude of statistical fluctuations in the two-time correlations $C$ for $l=4$, obtained according to the protocol of Sec.~\ref{s:protocol} from measurement samples of finite size. Real parts of the exact two-time correlation functions $C(0,t_2)$, starting from the uniform initial state \eqref{e:Psiini_uniform}, are plotted as orange solid lines. Blue dashed lines show estimates of $C$ obtained according to Eqs.~\eqref{e:ReC} and \eqref{e:wmcf} with the exact probabilities $P_\pm P_{m|\pm}$, which corresponds to using measurement samples of infinite size. The blue shaded areas around the blue dashed lines indicate statistical error bars of one standard deviation, calculated according to the procedure described in the text for measurement samples of finite size. For sample size $\mathscr{N}_\text{s}=100$ (left panel) statistical fluctuations are the dominant source of errors, larger than the systematic deviation between the estimated (blue dashed) and exact (orange solid) values. For sample size $\mathscr{N}_\text{s}=1000$ (right panel) statistical fluctuations are significantly reduced and are smaller than systematic errors.
}
\end{figure}

Figure \ref{f:errors} shows, for the uniform initial state \eqref{e:Psiini_uniform}, the $t_2$-dependence of the two-time correlation function $C(0,t_2)$ estimated according to the measurement protocol, together with the (likewise $t_2$-dependent) statistical errorbars obtained according to the procedure described in the previous paragraph. For sample size $\mathscr{N}_\text{s}=100$ the statistical errors are considerable (Fig.~\ref{f:errors} left), but already for $\mathscr{N}_\text{s}=1000$ statistical fluctuations are smaller than the systematic errors that result from the approximations in the measurement protocol of Sec.~\ref{s:protocol} (Fig.~\ref{f:errors} right). A rough estimate based on the available data suggests that statistical error bars of $C$ scale like $\mathscr{N}_\text{s}^{-1/2}$ with the sample size. Moreover, one finds, for fixed sample size $\mathscr{N}_\text{s}$ and at least for spin quantum numbers $l$ up to 32 for which we have data, that statistical errors are essentially independent of $l$ (not shown in the plots).

\section{Summary and conclusions}
\label{s:summary}

To experimentally determine two-time quantum correlation according to the measurement protocol proposed in this paper, the key task is to determine sufficiently accurate estimators of the probabilities $P_\pm P_{m|\pm}$ defined in Eqs.~\eqref{e:probt1} and \eqref{e:conditional_probt2}. These probabilities describe the likelihood of measuring $\pm1/2$ for the ancilla spin component $S^z$, followed by a measurement result $m$ for the component $S_j^z$ of the spin at site $j$. Estimators of the probabilities $P_\pm P_{m|\pm}$ can therefore be obtained as the relative frequencies of the outcome pairs $(\pm,m)$ recorded over a sufficiently large sample of measurements. Each run follows the measurement protocol of Sec.~\ref{s:protocol}, which consists of the following steps:
\begin{enumerate}
\itemsep-0.3mm
\renewcommand{\theenumi}{(\alph{enumi})}
\renewcommand{\labelenumi}{\theenumi}
\item Prepare the spin-$1/2$ ancilla in the state \eqref{e:ancilla_ini}.\label{i:firststep}
\item Time-evolve the system until time $t_1$.
\item Couple system and ancilla by means of the unitary \eqref{e:coupling}.
\item Measure the ancilla observable $S^z$.
\item Time-evolve the system until time $t_2$.
\item Measure the system observable $S_j^z$.\label{i:laststep}
\end{enumerate}
By recording the measurement outcomes and repeating the above steps multiple times, estimators of the probabilities \eqref{e:probt1} and \eqref{e:conditional_probt2} are obtained. From these probabilities the correlation function $\mathscr{C}$ as defined in \eqref{e:wmcf} can be computed. Repeating the above steps \ref{i:firststep}--\ref{i:laststep} for several couplings $\lambda$ then gives access to the real and imaginary parts of $\mathscr{C}$ separately, as discussed at the end of Sec.~\ref{s:CandC}.

The measured correlation function \eqref{e:wmcf} is, under suitable assumptions, related to the desired correlation function $C$ via Eq.~\eqref{e:C_final}. The assumptions that went into the calculation of this relation are as follows:
\begin{enumerate}
\itemsep-0.3mm
\renewcommand{\theenumi}{(\roman{enumi})}
\renewcommand{\labelenumi}{\theenumi}
\item\label{i:largel} Large spin quantum number $l\gg1$ of the spin at lattice site $i$. This is only really needed when going from \eqref{e:C_v6} to \eqref{e:C_v7} by assuming $\sqrt{1-(S_i^z/\mvec{S}_i)^2}\approx1$. This is a shorthand notation for the actual requirement that only those expectation values in the expansion \eqref{e:C_v5} contribute significantly to $\mathscr{C}$ for which $(\tilde m/l)^2\ll1$ and $(m/l)^2\ll1$. At other instances in the derivation, $l\gg1$ is only assumed for the convenience of shorter expressions; replacing occurrences of $l$ by $l+1/2$ in \eqref{e:C_final} will undo these further approximations.
\item\label{i:ass2} The expansion coefficients $\gamma_m^\pm$ \eqref{e:gamma} of the system state at $t_1$ vary slowly with $m$, such that $\gamma_{m+1}^\pm\approx\gamma_m^\pm$. 
\item\label{i:ass3} Assumption \eqref{e:matrix_approx} on the matrix elements of $S_j^z(t_2-t_1)$ is condition \ref{i:ass2} in disguise: the matrix elements are required to be slowly varying in $m$.
\end{enumerate}
While the largeness of the spin quantum number $l$ in condition \ref{i:largel} should be easy to assess, the slowly varying expansion coefficients $\gamma_m^\pm$ required in assumptions \ref{i:ass2} and \ref{i:ass3} may require additional experimental effort. In principle the coefficients $\gamma_m^\pm$ may be obtained in a separate series of experimental runs in which the spin component $S_i^z$ at site $i$ is measured at time $t_1$ and relative frequencies of the outcomes are recorded. (This may be referred to as full counting statistics \cite{Nazarov} or be viewed as a particularly simple instance of quantum state tomography \cite{ParisRehacek}.) Bear in mind, however, that full knowledge of the coefficients $\gamma_m^\pm$ is not required and there may be simpler ways of checking that the requirement of slow variation with $m$ is satisfied, for example by verifying that the corresponding Fourier series decays sufficiently fast.

An experimental platform in which an assembly of interacting large spins can be realized, and in which the protocol proposed in this paper can potentially be implemented, is discussed in Sec.~\ref{s:intro} and sketched in Fig.~\ref{f:System}. Other suitable platforms, based for example on multiply-degenerate groundstate manifolds, are expected to be feasible as well. Potential physical applications of the thus obtained two-time correlation functions include the analysis of temporal universality in the vicinity of quantum phase transitions, or of signatures of slow relaxation in quantum glasses. Since both of these mentioned applications are genuine many-body phenomena, I conclude by re-emphasizing that, even though Sec.~\ref{s:example} treats a simple two-spin system for illustrative purposes, the protocol put forward in this paper is suited for proper many-body applications and can in principle be applied to large-$l$ spin systems of arbitrary lattice dimension and lattice size. In comparison to measurement protocols based on weak system--ancilla couplings, such as those of Refs.\ \cite{Uhrich_etal17,KastnerUhrich18}, the strong-coupling protocol of the present paper has the advantage of requiring substantially smaller measurement samples in order to accumulate sufficient statistics for obtaining accurate estimators of the probabilities $P_\pm P_{m|\pm}$ of Eqs.~\eqref{e:probt1} and \eqref{e:conditional_probt2}.

\acknowledgments
Discussions with Markus Oberthaler, which initiated the study reported in this paper, are gratefully acknowledged.

\appendix
\section{System--ancilla coupling \texorpdfstring{$S_i^z\otimes S^z$}{Siz Sz}}
\label{a:zz}

The measurement protocol of Sec.~\ref{s:protocol} makes use of a system--ancilla coupling of Heisenberg type \eqref{e:coupling}. In this appendix I consider an alternative, namely a coupling unitary $\mathscr{U}(\lambda)=\exp(-i\lambda S_i^z\otimes S^z)$ of Ising type. Since the uncoupled basis states $\ket{l,m_l,s,m_s}$ are eigenstates of the system spin operator $S_i^z$ as well as of the ancilla spin operator $S^z$, no need for the use of a coupled basis arises. By expanding
\begin{equation}
\ket{\Psi(t_1)}=\sum_{m_l,m_s}c_{m_l m_s}\ket{l,m_l,s,m_s}
\end{equation}
in the uncoupled basis, one can write \eqref{e:wmcf} as
\begin{align}
\mathscr{C}=&\sum_{m_l,m_s}\sum_{\tilde m_l,\tilde m_s}c_{\tilde m_l \tilde m_s}^*c_{m_l m_s} \exp\left[i\lambda(\tilde m_l \tilde m_s - m_l m_s)\right]\nonumber\\
&\times\Braket{l,\tilde m_l,s,\tilde m_s|S_j^z(t_2-t_1)\otimes S^z|l,m_l,s,m_s}\nonumber\\
=&\tfrac{1}{2}\sum_{m_l,\tilde m_l}\Bigl(c_{\tilde m_l +}^* c_{m_l +} e^{i\lambda(\tilde m_l-m_l)}\Braket{l,\tilde m_l|S_j^z(t_2-t_1)|l,m_l}\nonumber\\
&-c_{\tilde m_l -}^* c_{m_l -} e^{-i\lambda(\tilde m_l-m_l)}\Braket{l,\tilde m_l|S_j^z(t_2-t_1)|l,m_l}\Bigr).
\end{align}
The form of the ancilla initial state \eqref{e:ancilla_ini} implies $c_{m_l +}=c_{m_l -}\equiv c_{m_l}$ for all $m_l$, which allows one to write
\begin{equation}\label{e:zz_final1}
\mathscr{C}=\sum_{m_l,\tilde m_l}c_{\tilde m_l}^* c_{m_l} \sin\left[\lambda(\tilde m_l-m_l)\right]\Braket{l,\tilde m_l|S_j^z(t_2-t_1)|l,m_l}
\end{equation}
or, by making use of the spectral theorem,
\begin{multline}
\mathscr{C}=\tfrac{1}{2}\Braket{\Psi(t_1)|e^{i\lambda S_i^z}S_j^z(t_2-t_1)e^{-i\lambda S_i^z}|\Psi(t_1)}\\
-\tfrac{1}{2}\Braket{\Psi(t_1)|e^{-i\lambda S_i^z}S_j^z(t_2-t_1)e^{i\lambda S_i^z}|\Psi(t_1)}.
\end{multline}
It is not evident how the two-time correlation function \eqref{e:Ct1t2} can be extracted from this quantity, except in the limit of small $\lambda$, which amounts to recovering one of the results of Ref.~\cite{Uhrich_etal17}.

The coupling unitary $\mathscr{U}(\lambda)=\exp(-i\lambda S_i^x\otimes S^x)$ leads to similar results, only with $\sin\left[\lambda(\tilde m_l-m_l)\right]$ replaced by $\cos\left[\lambda(\tilde m_l+m_l)\right]$ in Eq.~\eqref{e:zz_final1}.


\bibliography{../../../MK.bib}

\end{document}